# Correlative Analysis of Iron-Driven Structural, Optical, and Magnetic Properties in Natural Biotite Crystals

Raphaela de Oliveira[1*], Yara Galvão Gobato[2], Ronei C. de Oliveira[2], José R. de Toledo[2], Verônica C. Teixeira[1], Angelo Malachias[3], Cesar R. Rabahi[2], Chunwei Hsu[4], Adilson J. A. de Oliveira[2], Herre. S. J. van der Zant[4], Ingrid D. Barcelos[1] and Alisson R. Cadore[5,6*]

[1]Brazilian Synchrotron Light Laboratory (LNLS), Brazilian Center for Research in Energy and Materials (CNPEM), 13083-970, Campinas, Sao Paulo, Brazil
[2]Department of Physics, Federal University of São Carlos, 13565-905, São Carlos, São Paulo, Brazil
[3]Department of Physics, Federal University of Minas Gerais, 31270-901, Belo Horizonte, Minas Gerais, Brazil
[4]Kavli Institute of Nanoscience, Delft University of Technology, Lorentzweg 1, 2628 CJ Delft, The Netherlands
[5]Brazilian Nanotechnology National Laboratory (LNNano), Brazilian Center for Research in Energy and Materials (CNPEM), 13083-970, Campinas, São Paulo, Brazil
[6]Programa de Pós-Graduação em Física, Instituto de Física, Universidade Federal do Mato Grosso, 79070-900, Cuiabá, Mato Grosso, Brazil

*Corresponding author: raphaela.goncalves@lnls.com; alisson.cadore@lnnano.cnpem.br

**Abstract**
Biotite crystals are phyllosilicate trioctahedral micas with the general chemical formula $K(Mg,Fe)_3AlSi_3O_{10}(OH)_2$ that form a solid-solution series with iron-poor phlogopite and iron-rich annite endmembers. With a wide band gap energy and a layered structure with free surface charges, biotite nanosheets can be readily obtained by cleavage methods and used as dielectrics in nanodevice fabrication for the next generation of electronics and energy harvesting. Here, a comprehensive study of biotite samples with different iron concentrations and oxidation states is presented. Structural, optical, magneto-optical, and magnetic characterizations were performed using several experimental techniques, including state-of-the-art synchrotron-based techniques, to correlate the iron chemistry (content and oxidation state) with the macroscopic properties of both minerals. The study reveals a nanoscale-homogeneous Fe distribution via synchrotron X-ray fluorescence mapping, defect-mediated optical transitions modulated by $Fe^{3+}/Fe^{2+}$ ratios, and temperature-dependent magnetic transitions from paramagnetism to competing ferro-/antiferromagnetic interactions. Furthermore, the use of these biotite crystals as substrates for ultrathin heterostructures incorporating monolayer (ML) $MoSe_2$ is explored by magneto photoluminescence at cryogenic temperatures. The results show that the presence of iron impurities in different oxidation states significantly impacts the valley properties for ML-$MoSe_2$. Overall, these findings offer a comprehensive interpretation of the physical properties of bulk biotites in a correlative approach, serving as a robust reference for future studies aiming to explore biotites in their ultrathin form.

## Introduction

One possible route for the future scaling of electronics is the integration of two-dimensional materials (2DMs) with existing silicon technology through the fabrication of thin nanosheet devices (Lemme et al., 2022; Zhu et al., 2023). Motivated by the development of this future hybrid electronic approach, researchers have investigated the fundamental properties and applications of several types of layered materials (LMs), including metallic and semimetallic ones (Novoselov et al., 2005; Craco et al., 2025), superconductors and semiconductors (Novoselov et al., 2005; Duan et al., 2015; Choi et al., 2017), as well as insulating (Illarionov et al., 2020; Knobloch et al., 2021) and topological insulating (Kou et al., 2017; Culcer et al., 2020) LMs, most of which are synthesized by different routes. Although there are naturally occurring LMs (Frisenda et al., 2020), relatively little research has been carried out on natural insulating LMs compared to synthetic ones (Barcelos et al., 2023). The usual presence of defects and impurities (i.e., extra elements incorporated in the crystal lattice during the material formation) in naturally occurring LMs poses challenges for optimizing their properties for nanotechnology applications. On the other hand, they are abundant on Earth and easy to extract. Therefore, understanding defect engineering that enables their use in future nanotechnological applications could be the key to reducing overall costs (Fiori et al., 2014; Illarionov et al., 2020; de Oliveira et al., 2024b; Lopez-Richard et al., 2025).

Recently, there has been increasing interest in the physical properties of phyllosilicate minerals in their few-layer form (Barcelos et al., 2018; Mogg et al., 2019; Santos et al., 2019; Frisenda et al., 2020; Matković et al., 2021; Zou et al., 2021; Cadore et al., 2022; de Oliveira et al., 2023; Longuinhos et al., 2023; Wei et al., 2023; Haley et al., 2024; Mahapatra et al., 2024a; Costa Freitas et al., 2025; Feres et al., 2025; Kawahala et al., 2025; Pacakova et al., 2025). Phyllosilicates are naturally occurring LMs with a large band gap energy, making them suitable as dielectrics in ultrathin van der Waals heterostructures (vdWHs) (Barcelos et al., 2018; Janica et al., 2018; Gadelha et al., 2021; Nutting et al., 2021; Prando et al., 2021; Vasic et al., 2021; Barbosa et al., 2025). Moreover, defects and impurities in this material class allow free charges on the material surface to be transferred to another LM, modifying the overall optical and electrical properties of the 2DM-based devices (Mania et al., 2017; Mahapatra et al., 2022, 2023; de Oliveira et al., 2024a; Pramanik et al., 2024; Ames et al., 2025). Phyllosilicates are Earth-abundant minerals, easy to extract, easily handled, and air-stable, with the common presence of iron (Fe) impurities, from which magnetic proximity effects and magnetic properties can arise, even at the monolayer limit (Matković et al., 2021; Khan et al., 2023; Mahapatra et al., 2024b; Pacakova et al., 2025).

Biotite emerges as a phyllosilicate of interest within the framework of Earth-abundant insulating LM sources, due to the possibility of tuning its optical and magnetic properties by controlling the Fe content. Also known as black mica, biotite is a phyllosilicate within the trioctahedral mica group with the chemical formula $K(Mg,Fe)_3AlSi_3O_{10}(OH)_2$ (W. A. Deer, R. A. Howie, 1962; Maslova et al., 2004; Deer et al., 2013) (see Figs. 1a-c). Moreover, biotite forms a solid-solution series with the phyllosilicate phlogopite $(KMg_3(AlSi_3)O_{10}(OH)_2)$, which is a Fe-free phyllosilicate in principle, and annite $(KFe_3AlSi_3O_{10}(OH)_2)$, the ideal Fe-endmember (W. A. Deer, R. A. Howie, 1962). The general lamellar structure of biotites (Fig. 1c) is oriented along the *c*-axis formed by the intercalation of two tetrahedral silicon oxide layers (T), with one Al-for-Si substitution, with (Mg,Fe)-trioctahedral layers (Oc) and K between the layers (W. A. Deer, R. A. Howie, 1962; Frisenda et al., 2020), forming a T-Oc-T structure with a cationic interlayer. The Oc layer has three possible sites, two of which are equivalent and called M1(B) sites, in which the six-coordinated ions are surrounded by four O and two OH in a *cis*-configuration (Rausell-Colom et al., 1979). The non-equivalent site is in a *trans*-configuration, called M1(A) site (Rausell-Colom et al., 1979). Fe atoms are expected to be present in biotites in different valence states $Fe^{2+/3+}$ in the octahedral sites. The typical M1(A)-O atomic distances are smaller than the M1(B)-O distances. Thus, ions with smaller ionic radius, such as $Fe^{3+}$ ions, tend to occupy the M1(A) sites, while $Fe^{2+}$ ions tend to occupy M1(B) sites, typically with twice the abundance. However, the Fe-end member of the phlogopite-annite series is expected to have only $Fe^{2+}$ ions occupying all octahedral sites. Furthermore, $Fe^{3+}$ ions can also occupy the Al-tetrahedral sites (Tripathi et al., 1978; Rausell-Colom et al., 1979; Rancourt et al., 1992).

This work provides a fresh understanding of the physical properties of bulk biotites in relation to their Fe content in different oxidation states, using a correlative approach to serve as a robust reference for studies exploring few-layer biotites embedded in vdWHs and ultrathin devices. The present work investigates the structural, optical, and magnetic properties of biotite crystals to improve the understanding of how Fe concentration and oxidation state influence these properties. Due to the natural origin of the minerals studied in this work, structural defects or minor contributions from trace elements may also partially influence their fundamental properties. However, the focus was placed on the major contributions resulting from differences in iron content. An extensive correlative study was carried out on two biotite samples with phlogopite-like and annite-like behaviors using several experimental techniques. Finally, the use of these crystals as substrates for vdWHs incorporating monolayer (ML) MoSe$_2$ is explored by magneto photoluminescence (PL) at cryogenic temperatures. These findings suggest biotite as a promising LM for developing low-cost magneto-optical and vdWHs-based devices due to its environmental stability and natural abundance.

**Results and discussion**

To investigate the structural properties of the biotite samples, XRD measurements were performed on freshly milled mineral powders (Fig. 1d). For Rietveld refinement of the diffractograms, biotite (Brigatti and Davoli, 1990), phlogopite (Hendricks and Jefferson, 1939), and annite (Redhammer et al., 2000) were considered as the main phases. The experimental data were fitted using the MAUD software package (Lutterotti et al., 1997), yielding good agreement with respect to peak intensities and positions, as indicated by the difference plot (orange line) between the fit and the experimental data for each sample. The most common biotite polytype exhibits monoclinic symmetry with a C2/m space group (Ross et al., 1966). For the phlogopite-like powder, the best fit was obtained with a volumetric combination of 40% biotite and 60% phlogopite phases. In contrast, for the annite-like powder, the best fit required a volumetric combination of 55% biotite and 45% annite phases. The structural parameters derived from the XRD fitting are summarized in Table 1. These results indicate that, although both samples are biotite minerals, the incorporation of Fe significantly affects the phase composition. In the phlogopite-like sample, whose diffractogram adjustment showed a dominant phlogopite contribution and a minor biotite contribution (red line), Mg is expected to be more abundant than Fe in the octahedral sites. On the other hand, the diffractogram of the annite-like sample could not be accurately fitted using only the biotite phase. An almost equal contribution from biotite and annite (blue line) was required, suggesting that Fe occupies a larger fraction of the octahedral sites compared to Mg in this sample. As will be discussed later, the Fe ions occupying the octahedral sites in the annite-like sample are predominantly in the Fe$^{2+}$ oxidation state, consistent with annite´s composition.

Table 1 - Crystal lattice parameters of phlogopite and annite-like biotites retrieved from Rietveld refinement.

| biotite sample | phase | a (Å) | b (Å) | c (Å) | α=γ (º) | β (º) |
|---|---|---|---|---|---|---|
| phlogopite-like | biotite | 5.367(5) | 9.266(8) | 10.196(2) | 90 | 100.12(6) |
| | phlogopite | 5.311(1) | 9.034(1) | 10.097(2) | 90 | 96.02(2) |
| annite-like | biotite | 5.367(8) | 9.228(9) | 10.04(1) | 90 | 101.8 (1) |
| | annite | 5.391(6) | 9.335(4) | 10.308(3) | 90 | 100.4(1) |

The Raman spectra of the samples are shown in Fig. 1e. Both the phlogopite-like and annite-like samples display twelve characteristic Raman modes at approximately 95, 140, 180, 405, 555, 680, 715, 760, 910, 1090, 3600, and 3650 cm$^{-1}$, appearing as broadened peaks with unresolved shoulders. The peak positions are nearly identical between the two samples. This high degree of spectral correspondence indicates that the general vibrational modes are consistent with those of biotite and annite, despite the expected predominance of Fe$^{2+}$-Oc sites for the annite-like sample. The main differences observed in the Raman spectrum of the annite-like, when compared to the phlogopite-like one, include more intense and better-defined peaks in the 3500-3800 cm$^{-1}$ range, attributed to OH-stretching modes (Aspiotis et al., 2022), as well as a more intense mode at 680 cm$^{-1}$, in addition to two extra pronounced modes at 265 and 640 cm$^{-1}$. The spectra are in good agreement with previously reported Raman spectra of biotite (Loh, 1973; Šontevska et al., 2007; Singh and Singha, 2016; Ulian and Valdrè, 2023). The Oc-layer vibrations dominate the spectrum up to 600 cm$^{-1}$, followed by T ring modes up to 800 cm$^{-1}$ and T-stretching modes up to 1200 cm$^{-1}$ (Aspiotis et al., 2022). The peaks corresponding to OH-stretching modes are highly sensitive to the chemical composition of the Oc sites (Wang et al., 2015; Aspiotis et al., 2022; Khan et al., 2023). In pure annite phases, only one OH-stretching peak is typically expected, resulting from the ordered occupation of $Fe^{2+}$-$Fe^{2+}$-$Fe^{2+}$ in the Oc sites. However, the annite-like sample investigated here contains two main mineralogical phases, involving varying occupancies of Fe, Mg, and possibly Al as substitutional ions in the Oc sites. This compositional variation can lead to the splitting of OH-stretching peaks, as also discussed in reference (Aspiotis et al., 2022). In contrast, the phlogopite-like sample shows weaker OH-stretching features, consistent with a mixed contribution from less probable Oc configurations. This suggests that the phlogopite-like sample incorporates more $Fe^{3+}$ than the annite-like one, enabling a greater variety of possible structural arrangements.

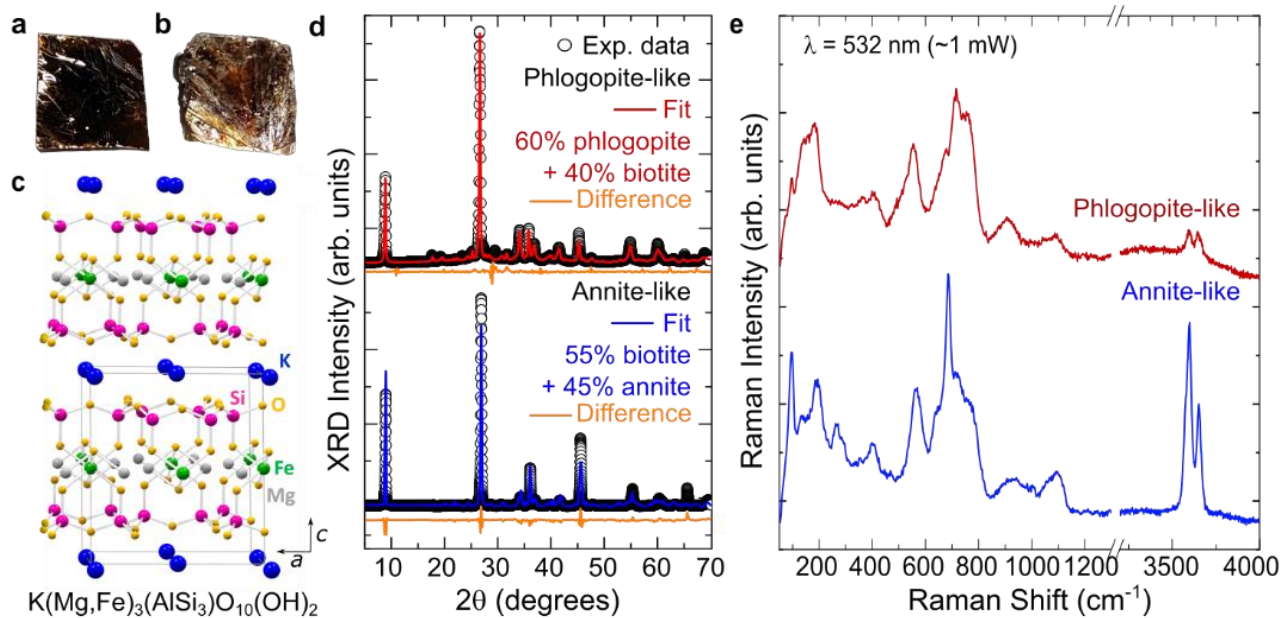

**Fig. 1.** Mineralogical characterization. a) Phlogopite-like and b) annite-like natural crystals with dimensions of ~ 5 x 5 x 1 mm$^3$. c) Atomic structure of Fe-rich biotites with general chemical formula K(Mg,Fe)$_3$(AlSi$_3$)O$_{10}$(OH)$_2$. d) XRD measurements (black circles) for phlogopite-like (top) and annite-like (bottom) samples fitted through Rietveld refinement considering contributions of phlogopite and biotite mineralogical phases for phlogopite-like sample (red) and biotite and annite-like phases for annite-like sample (blue), along with their difference (orange) from the experimental data. e) Raman spectra of phlogopite-like (red) and annite-like (blue) samples.

The chemical composition of the two samples was analyzed with particular focus on Fe incorporation. Representative EDS spectra for the phlogopite-like and annite-like samples are shown in Fig. 2a, normalized by the oxygen peak, which is the most intense. Based on the EDS data, the elemental composition of the samples can be qualitatively assessed. The spectra reveal the presence of aluminum (Al), potassium (K), magnesium (Mg), oxygen (O), and silicon (Si), and iron (Fe) as the main constituent elements. No additional impurities were detected. The observed sodium (Na) and carbon (C) signals are attributed to sample handling and environmental exposure. The primary spectral difference between the samples lies in the lower Mg content observed in the annite-like sample compared to the phlogopite-like one, which is compensated by higher concentrations of Al and Fe. This result is consistent with the XRD quantification, which indicated ~45% annite phase in the annite-like sample. Although EDS is a limited technique for quantitative analysis, it is possible to perform a semi-quantitative elemental analysis, taking care to choose a flat surface of freshly cleaved samples as representative regions, avoiding increasing error due to roughness and contamination. The semi-quantitative results are summarized in Table 2, presenting the concentrations of the main constituent elements (O, Si, Al, Fe, Mg, and K) from representative regions of each sample. The remaining detected content corresponds to Na and C. Our compositional analysis confirms that both phlogopite-like and annite-like samples exhibit major elemental concentrations in agreement with literature values (Zheng et al., 2020) (see Table 2). However, it is essential to note that the EDS has limited sensitivity for elements present at concentrations below ~2 wt%, which restricts the reliable quantification of trace impurities.

**Table 2** – Semi-quantitative analysis of phlogopite and annite-like biotites using EDS.

| Element | phlogopite-like (wt%) | annite-like (wt%) |
|---|---|---|
| O | 57 | 50 |
| Si | 15 | 18 |
| Al | 11 | 13 |
| Fe | 4 | 8 |
| Mg | 4 | 1 |
| K | 4 | 6 |

Now, to evaluate the oxidation states of iron in the biotite samples, spectroscopic characterizations were performed. UV-Vis-NIR absorption measurements, shown in Fig. 2b, revealed absorption bands associated with $Fe^{2+}$ and $Fe^{3+}$ valence states, as expected for phyllosilicate minerals (G. H. Faye, 1968). In the annite-like sample, the absorption band related to the $Fe^{2+}$-$Fe^{3+}$ charge exchange is almost suppressed, while the $Fe^{2+}$ bands are much more pronounced than in the phlogopite-like sample. This suggests that the Oc layer in the annite-like sample is chemically well-defined by a dominant presence of $Fe^{2+}$ ions. These findings are consistent with the EDS and Raman spectroscopy results, which indicate a more chemically diverse Oc layer in the phlogopite-like sample.

To further investigate the iron oxidation states, advanced synchrotron-based spectroscopic techniques were employed on multilayer exfoliated flakes. Using an X-ray nanoprobe, XRF and XANES measurements were conducted to probe the spatial distribution and oxidation state of Fe with sub-micrometric resolution. XRF microscopy was performed over a 50 x 50 μm² area on both phlogopite-like and annite-like flakes. Hyperspectral maps (insets in Figs. 2c,d) were generated by selecting the spectral response around the Fe-Kα emission energy, with a spatial resolution of 500 x 500 nm² per pixel. Both samples exhibit a homogeneous Fe distribution, with XRF intensity modulated by thickness (i.e., brighter areas = thicker sample regions, while darker areas = thinner sample regions). XANES spectra were acquired at the location marked by the white dots in the insets of Figs. 2c,d, and analyzed qualitatively by comparing them with reference spectra of iron oxides with known oxidation states (Figs. 2c,d). The phlogopite-like sample exhibits a magnetite-like signature, with its Fe-edge aligning with that of the $Fe_3O_4$ (magnetite) standard, indicating a mixed $Fe^{2+}$/$Fe^{3+}$ oxidation state. In contrast, the annite-like sample matches the absorption edge of the FeO standard, confirming the presence of iron predominantly in the $Fe^{2+}$ state. These results agree with our UV-Vis-NIR absorbance analysis, while also providing enhanced spatial resolution and sensitivity. The synchrotron-based X-ray nanoprobe enables the investigation of significantly smaller sample areas with sub-micrometric precision (Tolentino et al., 2023).

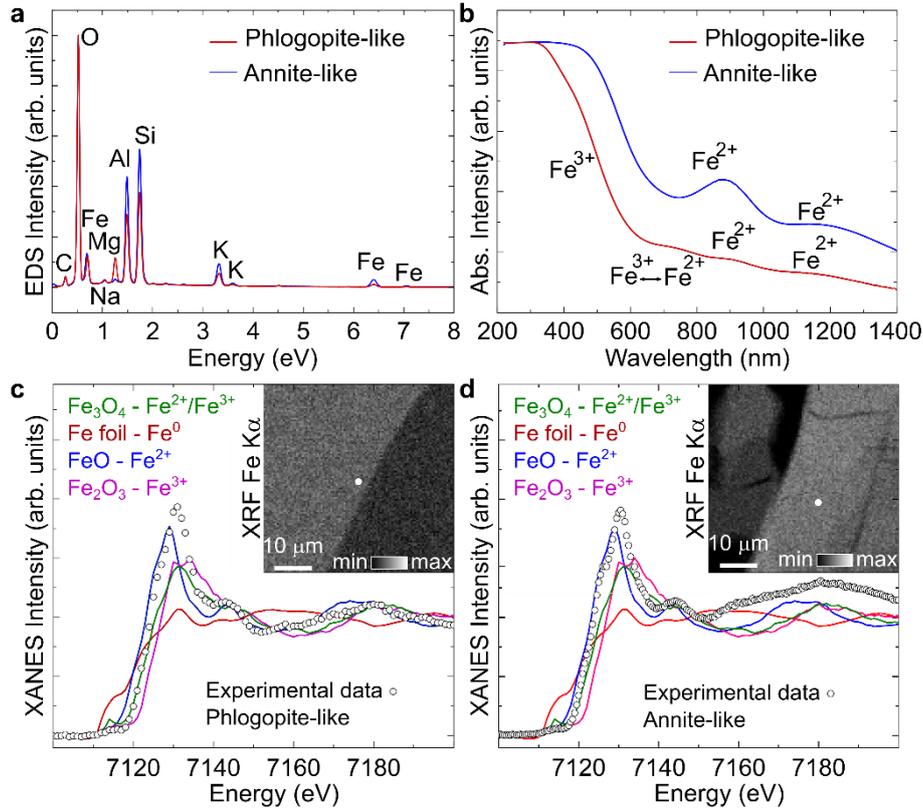

**Fig. 2.** Fe-content spectroscopy characterization. a) EDS spectra of phlogopite-like (red) and annite-like (blue) samples. b) UV-Vis-NIR absorbance spectra from the same samples. Fe K-edge XANES spectra collected at the location marked by a white dot in the corresponding XRF maps (insets) of c) phlogopite-like and d) annite-like samples, compared with reference iron oxides of known oxidation states. XRF mapping was performed with 500 nm spatial resolution at the Fe-Kα emission energy.

Next, the influence of Fe content on the optical properties of the samples was investigated. In particular, PL spectra were measured for both biotite crystals at various positions at 300 K, as shown in Figs. 3a-d. Overall, the PL spectra reveal different emission bands associated with optical transitions related to the presence of defects. For the biotite samples studied, these defects are mainly associated with Fe incorporation. Figs. 3a,b show typical PL spectra for both samples at 300 K. A broad, asymmetrical PL band centered ~2.8 eV is typically observed across most of the measurement positions. However, at certain locations, a stronger PL signal is detected with well-resolved emission bands, indicating multiple optical contributions (Figs. 3c,d). Deconvolution of the broad PL spectra by Voigt functions (represented by blue, green, purple, orange, and cyan areas) reveals the presence of four emission bands approximately 3.20, 3.00, 2.84, 2.71, and 2.65 eV for the phlogopite-like sample, and five bands at around 3.00, 2.84, 2.71, and 2.50 eV for the annite-like sample. The primary optical emissions from both samples are likely associated with transitions involving defect states with high energy, followed by relaxation processes resulting in lower-energy emissions. In general, the PL results are consistent with the reflectance data, and their origin is attributed to transitions related to Fe within the crystal structure.

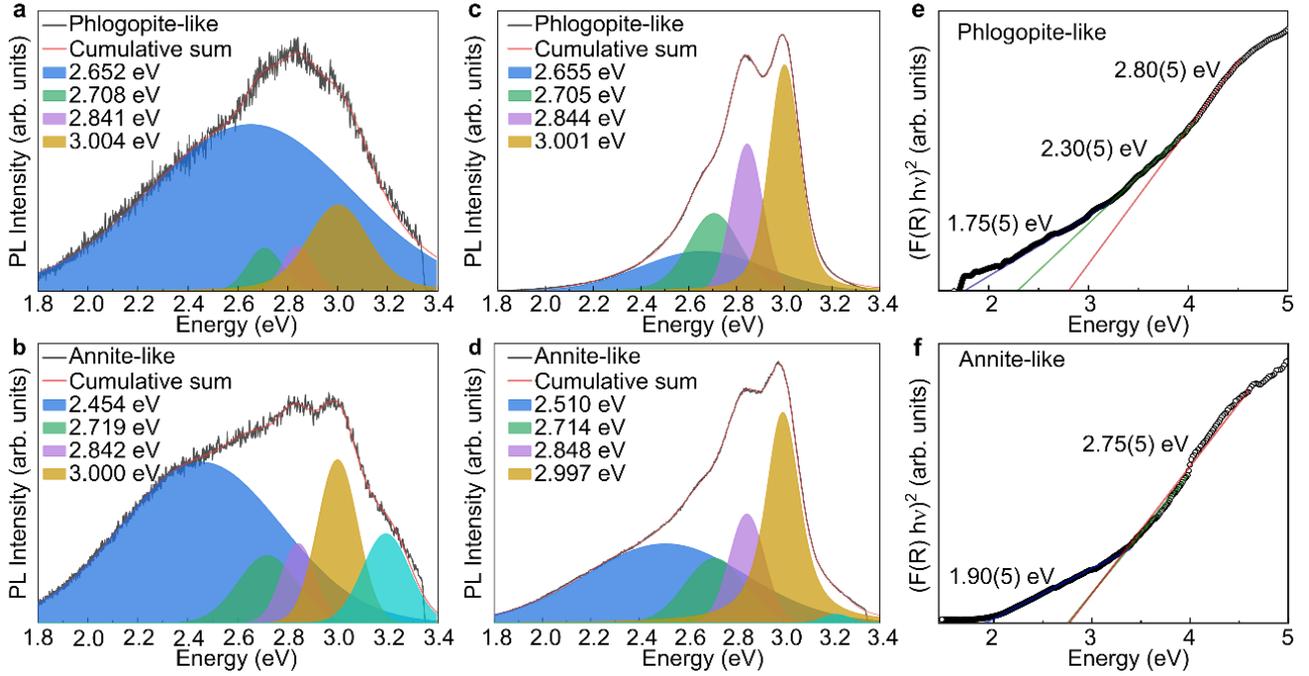

**Fig.3.** Optical properties of biotite samples. PL spectra acquired at 300 K for a) phlogopite-like and b) annite-like samples. PL obtained for measurement positions with stronger PL intensity for c) phlogopite-like and d) annite-like positions. The red lines are the cumulative sum of the individual emission peaks deconvolved in the blue, green, purple, orange, and cyan areas. Tauc plot obtained from UV-Vis-NIR reflectance measurements for e) phlogopite-like and f) annite-like samples showing the estimation of defect optical transitions (blue, green, and red lines).

To further investigate these transitions, UV-Vis-NIR reflectance measurements were carried out. For this, Tauc plots were derived from the reflectance spectra of the phlogopite-like (Fig. 3e) and annite-like (Fig. 3f) samples. The band gap energies were estimated by extrapolating the linear portions of the Kubelka–Munk function (Kubelka and Munk, 1931; Makuła et al., 2018). For the phlogopite-like sample, a nonlinear continuous trend is observed with increasing photon energy, suggesting a broad energy distribution of defect states. This behavior may arise from disordered defects caused by local site configuration variations or from defect band formation in regions with high defect concentration and long-range order. Both scenarios are plausible, given that biotite is inherently a solid solution with high Fe incorporation in various sites and oxidation states. Despite the continuous nature of the curve, three approximate linear regimes can be distinguished, corresponding to estimated optical transitions at 1.75(5) eV (blue line), 2.30(5) eV (green line), and 2.80(5) eV (red line). In contrast, the annite-like sample exhibits two well-defined linear regimes, separated by a discontinuity around 3 eV in Fig. 3f. The first regime yields an optical transition at 1.90(5) eV (blue line). In the second regime, a further discontinuity is observed near 4 eV, and two linear fits (green and red lines) converge to the same estimated optical transition at 2.75(5) eV. These different estimated band gaps indicate the presence of distinct defect levels involved in the optical transitions of the samples, which is consistent with the various emission bands observed in the PL spectra of the biotite samples (Fig. 3).

The magnetic properties of both phlogopite-like and annite-like biotite samples were thoroughly investigated through M(H) and M(T) measurements. The M(H) curves at 300 K and 2 K reveal a typical paramagnetic behavior for both samples, consistent with the presence of Fe in different oxidation states ($Fe^{2+}$ and $Fe^{3+}$) within the Oc layers. However, at 2 K, a small but noticeable hysteresis is observed, particularly in the phlogopite-like sample, suggesting the emergence of weak ferromagnetic interactions or spin-canting effects at low temperatures. This hysteresis is less pronounced in the annite-like sample, indicating a possible dependence of magnetic behavior on the Fe content and its distribution within the crystal lattice.

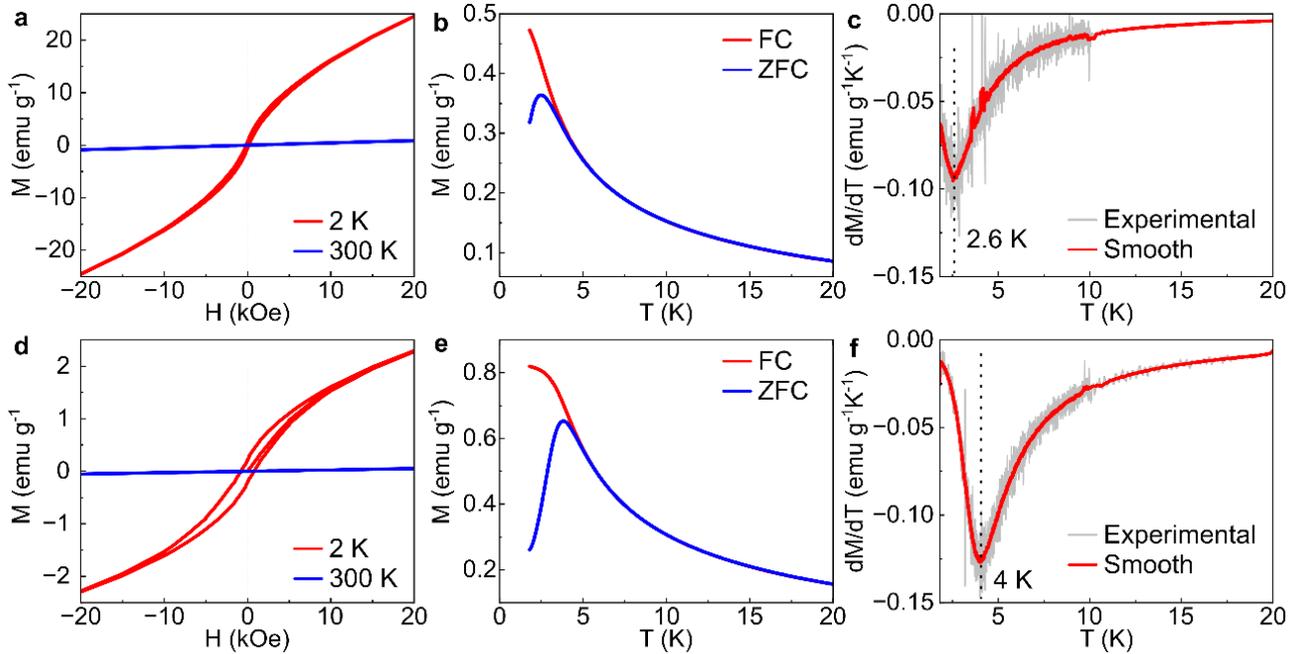

**Fig.4.** Magnetic characterization of biotite samples. a-c) Annite-like sample: a) M(H) hysteresis loop, b) FC-ZFC M(T) curve, and c) temperature derivative of magnetization (dM/dT) from FC measurements. d-f) Phlogopite-like sample: d) M(H) hysteresis loop, e) FC-ZFC M(T) curve, and f) dM/dT from FC measurements. All measurements were conducted with the magnetic field applied parallel to the sample surface, and the M(T) measurements were carried out under an applied magnetic field of 100 Oe.

The M(T) measurements further elucidate the magnetic properties of the samples. A distinct transition is observed in the ZFC curves, and the critical ordering temperatures were identified by analyzing the temperature derivative of magnetization (dM/dT) from the FC measurements. For the annite-like sample, the transition occurs at 2.6 K, whereas the phlogopite-like sample exhibits a transition at a slightly higher temperature of 4 K. These transitions may be attributed to spin freezing or the onset of short-range magnetic order, influenced by the differing iron concentrations and oxidation states in the two samples. The FC curves diverge from the ZFC curves below the transition temperatures, indicating magnetic irreversibility and the presence of competing magnetic interactions.

The observed differences in magnetic behavior between the two samples correlate with their structural and compositional characteristics, as revealed by XRD, EDS, XANES, Raman, and UV-Vis-NIR spectroscopy. The higher transition temperature and more prominent hysteresis in the phlogopite-like sample suggest a stronger influence of $Fe^{3+}$ ions, which may occupy specific Oc sites and contribute to localized magnetic moments. In contrast, the annite-like sample, with a higher $Fe^{2+}$ content, shows a lower transition temperature, reflecting distinct magnetic interactions associated with $Fe^{2+}$-dominated environments. These results are consistent with previous studies (Beausoleil et al., 1983; Khan et al., 2023), which indicate that a small fraction of $Fe^{3+}$ enhances ferromagnetic interactions when it is surrounded only by $Fe^{2+}$ ions.

However, as the $Fe^{3+}$ concentration increases, the formation of $Fe^{3+}$-$Fe^{3+}$ pairs becomes more likely, leading to dominant antiferromagnetic interactions. Thus, the present study provides experimental confirmation, demonstrating that although the total iron concentration significantly influences the mineral's properties, it is the $Fe^{3+}/Fe^{2+}$ ratio that serves as the key parameter for tuning its magnetic behavior. This finding may also be relevant for flakes with few-layer thicknesses and could be extended to explain recent results on vermiculite crystals (Pacakova et al., 2025).

Finally, the use of these crystals with different iron concentrations and oxidation states as ultrathin substrates for ML-MoSe$_2$ is explored. To this end, magneto-PL measurements were performed on ML-MoSe$_2$/phlogopite(annite)-like heterostructures (Fig. 5a,b) assembled onto SiO$_2$/Si substrates under out-of-plane magnetic field. Fig. 5a,b shows the typical $\sigma^+$ and $\sigma^-$ PL spectra recorded at 9T and 3.6 K. In these spectra, two distinct PL peaks are observed, corresponding to exciton (X) and trion (T) emissions (Glazov et al., 2024). Fig. 5c,d present the Zeeman splitting (ΔE), fitted using $\Delta E = E^{\sigma+} - E^{\sigma-} = g \mu_B B$, where $\mu_B$ = 58 μeV/T is the Bohr magneton, B is the out-plane magnetic field, and g is the valley g-factor. $E^{\sigma+}$ and $E^{\sigma-}$ represent the PL peak energies of the σ+ or σ− components for either the T or X emissions (de Oliveira et al., 2024a; Glazov et al., 2024).

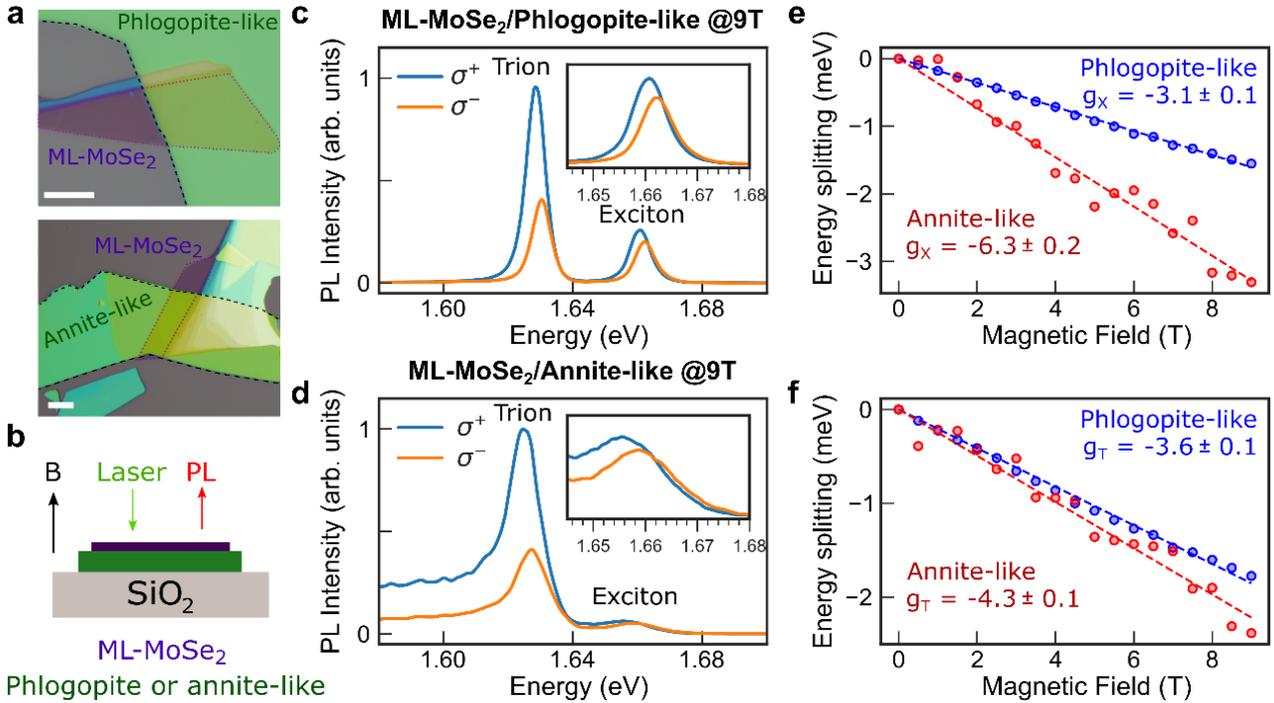

**Fig 5.** a) Optical microscope images of ML-MoSe$_2$ onto a phlogopite-like and annite-like flakes, assembled onto SiO$_2$/Si substrate. b) Schematic drawing of a vdWHs under an out-of-plane magnetic field. c,d) Circularly polarization resolved PL spectra (σ+ and σ-) for ML-MoSe$_2$/phlogopite-like and ML-MoSe$_2$/annite-like vdWHs, respectively, under out-of-plane magnetic fields at 3.6K. e,f) Energy splitting as a function of magnetic field for the exciton (X) and trion (T) emission peaks in ML-MoSe$_2$/phlogopite-like and ML-MoSe$_2$/annite-like vdWHs, respectively. The laser excitation is linearly polarized, and the PL detection is circularly polarized (σ+ and σ-).

The extracted X g-factor values are approximately −6.3 and −3.1 for the ML-MoSe$_2$ on annite- and phlogopite-like slabs, respectively. These results indicate a significant enhancement of the X g-factor for the ML-MoSe$_2$ on the annite-like flake, whereas a notable reduction is observed on the phlogopite-like substrate. The obtained g-factor values deviate considerably from typical values reported in literature, which generally range from −3.8 to −4.2 (Woźniak et al., 2020; Covre et al., 2022; Faria Junior et al., 2022; Gobato et al., 2022). This deviation is attributed to the presence of local iron impurities with varying oxidation states. The ML-MoSe$_2$ flake can hybridize with the underlying annite- and phlogopite-like layers, which may contain magnetic impurities at the interface (de Oliveira et al., 2024a). This hybridization could influence the valley properties of ML-MoSe$_2$. Furthermore, the extracted T g-factor values are approximately −4.3 and −3.6 for ML-MoSe$_2$ on annite- and phlogopite-like flakes, respectively. Interestingly, these values show only slight variations compared to those reported in the literature, suggesting that iron impurities predominantly affect the X g-factors rather than the T g-factors in ML-MoSe$_2$/biotite heterostructures, which is an unusual and intriguing result. Therefore, further studies are necessary to understand in detail the impact of iron impurities on the valley properties of ML-MoSe$_2$/phlogopite(annite)-like vdWHs.

## Conclusion

This work presents a comprehensive investigation of the structural, optical, and magnetic properties of two naturally occurring biotite samples with distinct iron contents and oxidation states. By employing a combination of advanced characterization techniques, including XRD, Raman spectroscopy, EDS, UV-Vis-NIR, XANES, PL, and SQUID magnetometry, we demonstrate that the Fe content and its valence state significantly influence the material's optical transitions and magnetic behavior. The phlogopite-like sample, richer in $Fe^{3+}$, shows enhanced magnetic ordering and broader optical transitions. In contrast, the annite-like sample, dominated by $Fe^{2+}$, exhibits more defined structural features and lower-temperature magnetic transitions. These findings establish a clear correlation between iron incorporation and the fundamental physical properties of biotite, providing critical insights for its use in future van der Waals heterostructures and magneto-optical devices.

In addition, it has been investigated the impact of using biotite crystals with different iron concentrations and oxidation states as magnetic substrates for MoSe$_2$ monolayers. The magneto-PL results show that the presence of iron impurities with varying oxidation states has a significant impact on the valley properties for ML-MoSe$_2$. In particular, it was observed unusual variations in the exciton g-factor values of ML-MoSe$_2$/phlogopite (annite)-like heterostructures. Furthermore, by providing experimental benchmarks on the oxidation states, atomic occupancy, and defect-sensitive properties of biotite, this work may serve as a valuable reference for theoretical studies. It can inspire first-principles simulations focused on the energetic stability of Fe ions in distinct lattice sites and oxidation states, as well as the role of defects and impurities in tuning the material's electronic and magnetic behavior at the atomic scale.

## Materials and Methods

This work investigates two geological biotite specimens obtained from locations in Brazil and England to understand compositional variations in naturally occurring biotite. Since biotite forms a solid-solution series spanning a wide range of Fe contents, two samples were selected: the Brazilian specimen, mineralogically closest to phlogopite (the Fe-free endmember), and the English specimen, closest to annite (the Fe-endmember). For clarity, they are referred to as phlogopite-like and annite-like samples, respectively.

*Mineralogical characterization:* X-ray diffraction (XRD) patterns were recorded in a Rigaku Miniflex II system with Cu-Kα radiation (λ = 1.5418 Å), employing a four-crystal Ge (220) (4-bounce) monochromator. The sample is measured in a two-axis diffractometer (operated in vertical geometry, in the Bragg–Brentano (θ:2θ) condition), using a NaI scintillator detector. The source divergence with this optical setup is nominally 1 mrad, and both receiving and anti-scatter slits were set to the minimum value of 0.625°. The generator was operated with a voltage and current of 30 kV and 15 mA, respectively. Raman spectra were acquired using a Horiba XploRA™ plus system with a 532 nm laser at 1 mW power. Spectra were accumulated 3 times for 30 s each, using a 1200 grooves/mm grating. Measurements employed a 100x objective (numerical aperture of 0.9), and the system was calibrated against the silicon peak at 521 cm$^{-1}$. Data were collected from ~1 μm diameter areas on mechanically exfoliated, bulk-like flat flakes transferred to silicon wafers.

*Iron-content characterization:* Energy-Dispersive Spectroscopy (EDS) was performed using a Thermo Scientific Helios 5 PFIB CXe DualBeam scanning electron microscope at an accelerating voltage of 15 kV and a current of 0.2 nA, in secondary electron imaging mode, with a chamber vacuum of 4.74 × 10$^{-5}$ Pa and an estimated spot size below 1 μm. Measurements were made on bulk biotite crystals (~5 x 5 mm$^2$ area) after the cleavage of superficial layers using mechanical exfoliation. The samples were glued using carbon tape in an aluminum sample holder. The EDS calibration was performed using a standard 100% silver (Ag) sample, resulting in a measurement error of approximately 2%. Multiple points were analyzed for statistical representation; data from a representative location reflecting average composition were selected, excluding anomalous regions. Absorbance and the reflectance spectra for Tauc plot analysis obtained using a Shimadzu 3600 Plus Ultraviolet–Visible–Near-Infrared (UV-Vis-NIR) spectrometer with 5.0 nm slit width. Measurements were conducted on bulk biotite crystals peeled off from the bulk crystals with homogeneous thickness (~5 x 5 mm$^2$ area) and fixed to the sample holder. X-ray fluorescence (XRF) hyperspectral maps of exfoliated biotite on Kapton tape were acquired at the CARNAUBA beamline (Tolentino et al., 2021) using synchrotron radiation over a 50 x 50 μm² area with 500 x 500 nm² pixel resolution. The X-ray beam spot measured 500 x 200 nm$^2$, with data collected in air at 9.75 keV excitation energy. Fe-Kα emission was selected for image generation. X-ray absorption near edge structure (XANES) spectra around the Fe-K edge were recorded in fluorescence mode at about 300 K. Standard references for Fe oxidation states (metallic Fe – Fe$^0$, FeO – Fe$^{2+}$, Fe$_2$O$_3$ – Fe$^{3+}$, and Fe$_3$O$_4$ – mixed Fe$^{2+}$, Fe$^{3+}$) were used for calibration and analyzed with Athena software (Ravel and Newville, 2005).

*Optical and magneto-optical characterization:* Photoluminescence (PL) measurements were performed in a custom setup optimized for the UV-visible range, featuring a Shamrock 500i spectrograph coupled with a CCD DU420A-BU detector and a continuous-wave 360 nm laser (CNI Model: UV-FN-360) at 1 mW. The estimated laser spot size on the sample is ~ 50 μm. For room-temperature measurements, bulk flat biotite fragments with ~5 x 5 mm$^2$ peeled from the bulk biotite crystals were mounted inside a closed-cycle Janis CCS-150 cryostat in vacuum (5 × 10$^{-6}$ mbar). Magneto-optical measurements were performed in a helium closed-cycle cryostat equipped with superconducting magnet coils (Attocube Attodry1000) and magnetic fields (B) of up to 9 T at 3.6 K. The vdWH samples were mounted on Attocube piezoelectric xyz translation stages to control the sample position. μ-PL measurements were performed using a continuous wave laser with a photon energy of 1.88 eV and 100 μW laser power with an estimated laser spot size on the sample of ~ 1 μm. The PL signal was collimated using an Attocube aspheric lens (NA = 0.64). The selection of circular polarization components (σ+ and σ-) was performed by using a linear polarizer and a quarter-wave plate before being focused into a 50 μm multimode optical fiber. The PL signal was then dispersed by a 75 cm Andor spectrometer and detected by a silicon CCD detector (Andor, Shamrock/iDus). Details can be found in Toledo et. al., (Roberto de Toledo et al., 2025).

*Magnetic characterization:* The magnetic properties of the phlogopite-like and annite-like biotite samples were investigated using a superconducting quantum interference device (SQUID) magnetometer (MPMS3, Quantum Design). Samples were prepared by cleaving a small fragment (~3 × 3 mm$^2$) from the bulk biotite crystals, and their masses were subsequently measured. Magnetization (M) measurements were conducted at 300 K and 2 K, with magnetic fields up to 20 kOe applied parallel to the sample surface. Additionally, temperature-dependent magnetization, M(T), was measured under zero-field-cooled (ZFC) and field-cooled (FC) protocols, using an applied magnetic field of 100 Oe. In the ZFC protocol, the samples were cooled down from 300 K to 2 K in zero-field. Once the temperature stabilized at 2 K, a magnetic field was applied, and data were collected during warming at a rate of 0.1 K/min. For the FC protocol, the samples were cooled under the applied field, and data were collected during cooling down. The temperature derivative of magnetization (dM/dT) was calculated to identify possible magnetic transitions.

*Van der Waals heterostructures stacking:* vdWHs composed of ML-MoSe$_2$ and phlogopite/annite-like flakes were prepared on SiO$_2$/Si substrates by mechanical exfoliation and the dry-transfer process technique. The ML-MoSe$_2$ flakes were obtained from the exfoliation of the MoSe$_2$-bulk crystal (2D Semiconductors) onto a polydimethylsiloxane (PDMS) stamp by the scotch tape technique and selected under an optical microscope (Castellanos-Gomez et al., 2014). The ML-MoSe$_2$/PDMS stamps were then aligned and stamped on the phlogopite (or annite)-like flakes slabs using a commercial transfer system (HQ Graphene).

**Acknowledgments**
All Brazilian authors thank the financial support from "Fundação Coordenação de Aperfeiçoamento de Pessoal de Nível Superior" (CAPES), "Conselho Nacional de Desenvolvimento Científico e Tecnológico" (CNPq), "Fundação de Amparo à Pesquisa do Estado de Minas Gerais" (FAPEMIG). The authors thank the LNNano (Proposals: FTQ-20221810, DRX-20230142, Raman-20230143, MNF-20230017, and MNF-20240039) and the Microscopic Samples Laboratory (LAM, at the LNLS – Proposals: LAM-2D-20230170, LAM-2D-20230164 and LAM-PFIB-20242188), and the CARNAUBA beamline at LNLS (Proposal: 20222162) both part of the CNPEM, a private nonprofit organization under the supervision of the Brazilian Ministry for Science, Technology, and Innovations (MCTI), for sample fabrication and characterization. This work was supported by "Fundação de Amparo à Pesquisa do Estado de São Paulo" (FAPESP) (grant numbers 21/10293-1, 22/00419-0, 22/08329-0, 23/11265-7, and 23/01313-4), and through the Research, Innovation and Dissemination Center for Molecular Engineering for Advanced Materials – CEMol (Grant CEPID No. 2024/00989-7), and Financiadora de Estudos e Projetos (FINEP) (grant 01.22.0178.00 – MATSemBGL (14796)). Finally, all authors thank Professor Marco A. Fonseca from Federal University of Ouro Preto for supplying the phlogopite-like biotite crystal, Dr. Jessica Fonsaca for enlightening discussion, and Carolina Torres, Cilene Labre, and Karen Bedin for experimental assistance.

**References**

Ames, A., Sousa, F.B., Souza, G.A.D., de Oliveira, R., Silva, I.R.F., Rodrigues, G.L., Watanabe, K., Taniguchi, T., Marques, G.E., Barcelos, I.D., Cadore, A.R., Lopez-Richard, V., Teodoro, M.D., 2025. Optical Memory in a MoSe2/Clinochlore Device. ACS Appl Mater Interfaces 17, 12818–12826. https://doi.org/10.1021/acsami.4c19337

Aspiotis, S., Schlüter, J., Redhammer, G.J., Mihailova, B., 2022. Non-destructive determination of the biotite crystal chemistry using Raman spectroscopy: how far we can go? European Journal of Mineralogy 34, 573–590. https://doi.org/10.5194/ejm-34-573-2022

Barbosa, T.C., Chaves, A.J., Freitas, R.O., Campos, L.C., Barcelos, I.D., 2025. Ultra-confined plasmons reveal moiré patterns in a twisted bilayer graphene–talc heterostructure. Nanoscale 17, 9205–9212. https://doi.org/10.1039/D4NR04532G

Barcelos, I.D., Cadore, A.R., Alencar, A.B., Maia, F.C.B., Mania, E., Oliveira, R.F., Bufon, C.C.B., Malachias, Â., Freitas, R.O., Moreira, R.L., Chacham, H., 2018. Infrared Fingerprints of Natural 2D Talc and Plasmon–Phonon Coupling in Graphene–Talc Heterostructures. ACS Photonics 5, 1912–1918. https://doi.org/10.1021/acsphotonics.7b01017

Barcelos, I.D., de Oliveira, R., Schleder, G.R., Matos, M.J.S., Longuinhos, R., Ribeiro-Soares, J., Barboza, A.P.M., Prado, M.C., Pinto, E.S., Gobato, Y.G., Chacham, H., Neves, B.R.A., Cadore, A.R., 2023. Phyllosilicates as earth-abundant layered materials for electronics and optoelectronics: Prospects and challenges in their ultrathin limit. J Appl Phys 134, 090902. https://doi.org/10.1063/5.0161736

Beausoleil, N., Lavallée, P., Yelon, A., Ballet, O., Coey, J.M.D., 1983. Magnetic properties of biotite micas. J Appl Phys 54, 906–915. https://doi.org/10.1063/1.332053

Brigatti, M.F., Davoli, P., 1990. Crystal-structure refinements of 1M plutonic biotites. American Mineralogist 75, 305–313.

Cadore, A.R., de Oliveira, R., Longuinhos, R., Teixeira, V. de C., Nagaoka, D.A., Alvarenga, V.T., Ribeiro-Soares, J., Watanabe, K., Taniguchi, T., Paniago, R.M., Malachias, A., Krambrock, K., Barcelos, I.D., de Matos, C.J.S., 2022. Exploring the structural and optoelectronic properties of natural insulating phlogopite in van der Waals heterostructures. 2d Mater 9, 035007. https://doi.org/10.1088/2053-1583/ac6cf4

Castellanos-Gomez, A., Buscema, M., Molenaar, R., Singh, V., Janssen, L., van der Zant, H.S.J., Steele, G. a, 2014. Deterministic transfer of two-dimensional materials by all-dry viscoelastic stamping. 2d Mater 1, 011002. https://doi.org/10.1088/2053-1583/1/1/011002

Choi, W., Choudhary, N., Han, G.H., Park, J., Akinwande, D., Lee, Y.H., 2017. Recent development of two-dimensional transition metal dichalcogenides and their applications. Materials Today 20, 116–130. https://doi.org/10.1016/j.mattod.2016.10.002

Costa Freitas, L.V., Rodrigues-Junior, G., Bernardes Marçal, L.A., Calligaris, G.A., Barcelos, I.D., de Oliveira, R., Malachias, A., 2025. Comparative study of phyllosilicate surface and sub-surface interlamellar water adsorption through crystal truncation rod modelling. Appl Surf Sci 682, 161775. https://doi.org/10.1016/j.apsusc.2024.161775

Covre, F.S., Faria, P.E., Gordo, V.O., de Brito, C.S., Zhumagulov, Y. V., Teodoro, M.D., Couto, O.D.D., Misoguti, L., Pratavieira, S., Andrade, M.B., Christianen, P.C.M., Fabian, J., Withers, F., Galvão Gobato, Y., 2022. Revealing the impact of strain in the optical properties of bubbles in monolayer MoSe2. Nanoscale 14, 5758–5768. https://doi.org/10.1039/D2NR00315E

Craco, L., Carara, S.S., Chagas, E.F., Cadore, A.R., Leoni, S., 2025. Electronic correlation and Mott localization of paramagnetic CrSBr crystal. Eur Phys J B 98, 145. https://doi.org/10.1140/epjb/s10051-025-00991-6

Culcer, D., Cem Keser, A., Li, Y., Tkachov, G., 2020. Transport in two-dimensional topological materials: recent developments in experiment and theory. 2d Mater 7, 022007. https://doi.org/10.1088/2053-1583/ab6ff7

de Oliveira, R., Barbosa Yoshida, A.B., Rabahi, C.R., O Freitas, R., Teixeira, V.C., de Matos, C.J.S., Galvão Gobato, Y., Barcelos, I.D., Cadore, A.R., 2024a. Ultrathin natural biotite crystals as a dielectric layer for van der Waals heterostructure applications. Nanotechnology 35, 505703. https://doi.org/10.1088/1361-6528/ad7b3a

de Oliveira, R., Cadore, A.R., Freitas, R.O., Barcelos, I.D., 2023. Review on infrared nanospectroscopy of natural 2D phyllosilicates. Journal of the Optical Society of America A 40, C157–C168. https://doi.org/10.1364/JOSAA.482518

de Oliveira, R., Freitas, L.V.C., Chacham, H., Freitas, R.O., Moreira, R.L., Chen, H., Hammarberg, S., Wallentin, J., Rodrigues-Junior, G., Marçal, L.A.B., Calligaris, G.A., Cadore, A.R., Krambrock, K., Barcelos, I.D., Malachias, A., 2024b. Water Nanochannels in Ultrathin Clinochlore Phyllosilicate Mineral with Ice-like Behavior. Journal of Physical Chemistry C 128, 14388–14398. https://doi.org/10.1021/acs.jpcc.4c02170

Deer, W.A., Howie, R.A., Zussman, J., 2013. An Introduction to the Rock-Forming Minerals. Mineralogical Society of Great Britain and Ireland. https://doi.org/10.1180/DHZ

Duan, Xidong, Wang, C., Pan, A., Yu, R., Duan, Xiangfeng, 2015. Two-dimensional transition metal dichalcogenides as atomically thin semiconductors : opportunities and challenges. Chem Soc Rev 44, 8859–8876. https://doi.org/10.1039/C5CS00507H

Faria Junior, P.E., Zollner, K., Woźniak, T., Kurpas, M., Gmitra, M., Fabian, J., 2022. First-principles insights into the spin-valley physics of strained transition metal dichalcogenides monolayers. New J Phys 24, 083004. https://doi.org/10.1088/1367-2630/ac7e21

Feres, F.H., Maia, F.C.B., Chen, S., Mayer, R.A., Obst, M., Hatem, O., Wehmeier, L., Nörenberg, T., Queiroz, M.S., Mazzotti, V., Klopf, J.M., Kehr, S.C., Eng, L.M., Cadore, A.R., Hillenbrand, R., Freitas, R.O., Barcelos, I.D., 2025. Two-dimensional talc as a natural hyperbolic material.

Fiori, G., Bonaccorso, F., Iannaccone, G., Palacios, T., Neumaier, D., Seabaugh, A., Banerjee, S.K., Colombo, L., 2014. Electronics based on two-dimensional materials. Nat Nanotechnol 9, 768–779. https://doi.org/10.1038/nnano.2014.207

Frisenda, R., Niu, Y., Gant, P., Muñoz, M., Castellanos-Gomez, A., 2020. Naturally occurring van der Waals materials. NPJ 2D Mater Appl 4, 1–13. https://doi.org/10.1038/s41699-020-00172-2

G. H. Faye, 1968. The optical absorption spectra of iron in six-coordinate sites in chlorite, biotite, phlogopite and vivianite; some aspects of pleochroism in the sheet silicates. Can Mineral 9, 403–425.

Gadelha, A.C., Vasconcelos, T.L., Cançado, L.G., Jorio, A., 2021. Nano-optical Imaging of In-Plane Homojunctions in Graphene and MoS 2 van der Waals Heterostructures on Talc and SiO 2. J Phys Chem Lett 12, 7625–7631. https://doi.org/10.1021/acs.jpclett.1c01804

Glazov, M., Arora, A., Chaves, A., Gobato, Y.G., 2024. Excitons in two-dimensional materials and heterostructures: Optical and magneto-optical properties. MRS Bull 49, 899–913. https://doi.org/10.1557/s43577-024-00754-1

Gobato, Y.G., de Brito, C.S., Chaves, A., Prosnikov, M.A., Woźniak, T., Guo, S., Barcelos, I.D., Milošević, M. V., Withers, F., Christianen, P.C.M., 2022. Distinctive g-Factor of Moiré-Confined Excitons in van der Waals Heterostructures. Nano Lett 22, 8641–8646. https://doi.org/10.1021/acs.nanolett.2c03008

Haley, K.L., Lee, N.F., Schreiber, V.M., Pereira, N.T., Sterbentz, R.M., Chung, T.Y., Island, J.O., 2024. Isolation and Characterization of Atomically Thin Mica Phyllosilicates. ACS Appl Nano Mater 7, 25233–25240. https://doi.org/10.1021/acsanm.4c05338

Hendricks, S.B., Jefferson, M.E., 1939. Polymorphism of the micas with optical measurements. Journal of the Mineralogical Society of America 24, 729–771.

Illarionov, Y.Y., Knobloch, T., Jech, M., Lanza, M., Akinwande, D., Vexler, M.I., Mueller, T., Lemme, M.C., Fiori, G., Schwierz, F., Grasser, T., 2020. Insulators for 2D nanoelectronics: the gap to bridge. Nat Commun 11, 3385. https://doi.org/10.1038/s41467-020-16640-8

Janica, I., Del Buffa, S., Mikołajczak, A., Eredia, M., Pakulski, D., Ciesielski, A., Samorì, P., 2018. Thermal insulation with 2D materials: Liquid phase exfoliated vermiculite functional nanosheets. Nanoscale 10, 23182–23190. https://doi.org/10.1039/c8nr08364a

Kawahala, N.M., Matos, D.A., de Oliveira, R., Longuinhos, R., Ribeiro-Soares, J., Barcelos, I.D., Hernandez, F.G.G., 2025. Shaping terahertz waves using anisotropic shear modes in a van der Waals mineral. NPJ 2D Mater Appl 9, 16. https://doi.org/10.1038/s41699-025-00540-w

Khan, M.Z., Peil, O.E., Sharma, A., Selyshchev, O., Valencia, S., Kronast, F., Zimmermann, M., Aslam, M.A., Raith, J.G., Teichert, C., Zahn, D.R.T., Salvan, G., Matković, A., 2023. Probing Magnetic Ordering in Air Stable Iron-Rich Van der Waals Minerals. Advanced Physics Research 2300070, 1–13. https://doi.org/10.1002/apxr.202300070

Knobloch, T., Illarionov, Y.Y., Ducry, F., Schleich, C., Wachter, S., Watanabe, K., Taniguchi, T., Mueller, T., Waltl, M., Lanza, M., Vexler, M.I., Luisier, M., Grasser, T., 2021. The performance limits of hexagonal boron nitride as an insulator for scaled CMOS devices based on two-dimensional materials. Nat Electron 4, 98–108. https://doi.org/10.1038/s41928-020-00529-x

Kou, L., Ma, Y., Sun, Z., Heine, T., Chen, C., 2017. Two-Dimensional Topological Insulators: Progress and Prospects. J Phys Chem Lett 8, 1905–1919. https://doi.org/10.1021/acs.jpclett.7b00222

Kubelka, P., Munk, F., 1931. An article on optics of paint layers. Zeitschrift für Physik 12, 259–274.

Lemme, M.C., Akinwande, D., Huyghebaert, C., Stampfer, C., 2022. 2D materials for future heterogeneous electronics. Nat Commun. 13, 1392. https://doi.org/10.1038/s41467-022-29001-4

Loh, E., 1973. Optical vibrations in sheet silicates. Journal of Physics C: Solid State Physics 6, 1091–1104. https://doi.org/10.1088/0022-3719/6/6/022

Longuinhos, R., Cadore, A.R., Bechtel, H.A., J S De Matos, C., Freitas, R.O., Ribeiro-Soares, J., Barcelos, I.D., 2023. Raman and Far-Infrared

Synchrotron Nanospectroscopy of Layered Crystalline Talc: Vibrational Properties, Interlayer Coupling, and Symmetry Crossover. The Journal of Physical Chemistry C 127, 5876–5885. https://doi.org/10.1021/acs.jpcc.3c00017

Lopez-Richard, V., Filgueira e Silva, I.R., Ames, A., Sousa, F.B., Teodoro, M.D., Barcelos, I.D., de Oliveira, R., Cadore, A.R., 2025. The Emergence of Mem-Emitters. Nano Lett 25, 1816–1822. https://doi.org/10.1021/acs.nanolett.4c04586

Lutterotti, L., Matthies, S., Wenk, H.-R., Schultz, A.S., Richardson, J.W., 1997. Combined texture and structure analysis of deformed limestone from time-of-flight neutron diffraction spectra. J Appl Phys 81, 594–600. https://doi.org/10.1063/1.364220

Mahapatra, P.L., Costin, G., Galvao, D.S., Lahiri, B., Glavin, N., Roy, A.K., Ajayan, P.M., Tiwary, C.S., 2024a. A comprehensive review of atomically thin silicates and their applications. 2d Mater 11, 032003. https://doi.org/10.1088/2053-1583/ad569b

Mahapatra, P.L., de Oliveira, C.C., Costin, G., Sarkar, S., Autreto, P.A.S., Tiwary, C.S., 2024b. Paramagnetic two-dimensional silicon-oxide from natural silicates. 2d Mater 11, 015019. https://doi.org/10.1088/2053-1583/ad10b9

Mahapatra, P.L., Singh, A.K., Tromer, R., R., K., M., A., Costin, G., Lahiri, B., Kundu, T.K., Ajayan, P.M., Altman, E.I., Galvao, D.S., Tiwary, C.S., 2023. Energy harvesting using two-dimensional (2D) d-silicates from abundant natural minerals. J Mater Chem C Mater 11, 2098–2106. https://doi.org/10.1039/D2TC04605A

Mahapatra, P.L., Tromer, R., Pandey, P., Costin, G., Lahiri, B., Chattopadhyay, K., P. M., A., Roy, A.K., Galvao, D.S., Kumbhakar, P., Tiwary, C.S., 2022. Synthesis and Characterization of Biotene: A New 2D Natural Oxide From Biotite. Small 18, 2201667. https://doi.org/10.1002/smll.202201667

Makuła, P., Pacia, M., Macyk, W., 2018. How To Correctly Determine the Band Gap Energy of Modified Semiconductor Photocatalysts Based on UV–Vis Spectra. J Phys Chem Lett 9, 6814–6817. https://doi.org/10.1021/acs.jpclett.8b02892

Mania, E., Alencar, A.B., Cadore, A.R., Carvalho, B.R., Watanabe, K., Taniguchi, T., Neves, B.R.A., Chacham, H., Campos, L.C., 2017. Spontaneous doping on high quality talc-graphene-hBN van der Waals heterostructures. 2d Mater 4, 031008. https://doi.org/10.1088/2053-1583/aa76f4

Maslova, M. v., Gerasimova, L.G., Forsling, W., 2004. Surface properties of cleaved mica. Colloid Journal 66, 364–371. https://doi.org/10.1023/B:COLL.0000030843.30563.c9

Matković, A., Ludescher, L., Peil, O.E., Sharma, A., Gradwohl, K.-P., Kratzer, M., Zimmermann, M., Genser, J., Knez, D., Fisslthaler, E., Gammer, C., Lugstein, A., Bakker, R.J., Romaner, L., Zahn, D.R.T., Hofer, F., Salvan, G., Raith, J.G., Teichert, C., 2021a. Iron-rich talc as air-stable platform for magnetic two-dimensional materials. NPJ 2D Mater Appl 5, 94. https://doi.org/10.1038/s41699-021-00276-3

Mogg, L., Hao, G.P., Zhang, S., Bacaksiz, C., Zou, Y.C., Haigh, S.J., Peeters, F.M., Geim, A.K., Lozada-Hidalgo, M., 2019. Atomically thin micas as proton-conducting membranes. Nat Nanotechnol. https://doi.org/10.1038/s41565-019-0536-5

Novoselov, K.S., Jiang, D., Schedin, F., Booth, T.J., Khotkevich, V. V, Morozov, S. V, Geim, A.K., 2005. Two-dimensional atomic crystals. Proc Natl Acad Sci U S A 102, 10451–10453. https://doi.org/10.1073/pnas.0502848102

Nutting, D., Prando, G.A., Severijnen, M., Barcelos, I.D., Guo, S., Christianen, P.C.M., Zeitler, U., Galvão Gobato, Y., Withers, F., 2021. Electrical and optical properties of transition metal dichalcogenides on talc dielectrics. Nanoscale 13, 15853–15858. https://doi.org/10.1039/D1NR04723J

Pacakova, B., Lahtinen-Dahl, B., Kirch, A., Demchenko, H., Osmundsen, V., Fuller, C.A., Chernyshov, D., Zakutna, D., Miranda, C.R., Raaen, S., Fossum, J.O., 2025. Naturally occurring 2D semiconductor with antiferromagnetic ground state. NPJ 2D Mater Appl 9, 38. https://doi.org/10.1038/s41699-025-00561-5

Pramanik, A., Mahapatra, P.L., Tromer, R., Xu, J., Costin, G., Li, C., Saju, S., Alhashim, S., Pandey, K., Srivastava, A., Vajtai, R., Galvao, D.S., Tiwary, C.S., Ajayan, P.M., 2024. Biotene: Earth-Abundant 2D Material as Sustainable Anode for Li/Na-Ion Battery. ACS Appl Mater Interfaces 16, 2417–2427. https://doi.org/10.1021/acsami.3c15664

Prando, G.A., Severijnen, M.E., Barcelos, I.D., Zeitler, U., Christianen, P.C.M., Withers, F., Galvão Gobato, Y., 2021. Revealing Excitonic Complexes in Monolayer WS2 on Talc Dielectric. Phys Rev

Appl 16, 64055. https://doi.org/10.1103/PhysRevApplied.16.064055

Rancourt, D.G., Dang, M.Z., Lalonde, A.E., 1992. Mössbauer spectroscopy of tetrahedral Fe3+ in trioctahedral micas. American Mineralogist 77, 34–43.

Rausell-Colom, J.A., Sanz, J., Fernandez, M., Serratosa, J.M., 1979. Distribution of Octahedral Ions in Phlogopites and Biotites, in: Mortaland, M.M., Farmer, V.C. (Eds.), Developments in Sedimentology,. Elsevier, Oxford, pp. 27–36. https://doi.org/10.1016/S0070-4571(08)70698-8

Ravel, B., Newville, M., 2005. ATHENA, ARTEMIS, HEPHAESTUS: data analysis for X-ray absorption spectroscopy using IFEFFIT. J Synchrotron Radiat 12, 537–541. https://doi.org/10.1107/S0909049505012719

Redhammer, G.J., Beran, A., Schneider, J., Amthauer, G., Lottermoser, W., 2000. Spectroscopic and structural properties of synthetic micas on the annite-siderophyllite binary: Synthesis, crystal structure refinement, Mössbauer, and infrared spectroscopy. American Mineralogist 85, 449–465. https://doi.org/10.2138/am-2000-0406

Roberto de Toledo, J., Serati de Brito, C., Rosa, B.L.T., Cadore, A.R., Rabahi, C.R., Faria Junior, P.E., Ferreira de Brito, A.C., Ghiasi, T.S., Ingla-Aynés, J., Schüller, C., van der Zant, H.S.J., Reitzenstein, S., Barcelos, I.D., Dirnberger, F., Gobato, Y.G., 2025. Interplay of Energy and Charge Transfer in WSe2/CrSBr Heterostructures. Nano Lett 25, 13212–13220. https://doi.org/10.1021/acs.nanolett.5c03150

Ross, M., Takeda, H., Wones, D.R., 1966. Mica Polytypes: Systematic Description and Identification. Science 151, 191–193.

Santos, J.C.C., Barboza, A.P.M., Matos, M.J.S., Barcelos, I.D., Fernandes, T.F.D., Soares, E.A., Moreira, R.L., Neves, B.R.A., 2019. Exfoliation and characterization of a two-dimensional serpentine-based material. Nanotechnology 30, 445705. https://doi.org/10.1088/1361-6528/ab3732

Singh, L., Singha, M., 2016. Vibrational spectroscopic study of muscovite and biotite layered phyllosilicates. Indian Journal of Pure & Applied Physics 54, 116–122.

Šontevska, V., Jovanovski, G., Makreski, P., 2007. Minerals from Macedonia. Part XIX. Vibrational spectroscopy as identificational tool for some sheet silicate minerals. J Mol Struct 834–836, 318–327. https://doi.org/10.1016/j.molstruc.2006.10.026

Tolentino, H., Geraldes, R.R., Moreno, G.B.Z.L., Pinto, A.C., Bueno, C.S., Kofukuda, L.M., Sotero, A.P., Neto, A.C., Lena, F.R., Wilendorf, W.H., Baraldi, G.L., Luiz, S.A., Dias, C.S.B., Perez, C.A., Neckel, I.T., Galante, D., Teixeira, V.C., Hesterberg, D., 2021. X-ray microscopy developments at Sirius-LNLS: first commissioning experiments at the Carnauba beamline, in: Lai, B., Somogyi, A. (Eds.), X-Ray Nanoimaging: Instruments and Methods V. SPIE, p. 6. https://doi.org/10.1117/12.2596496

Tolentino, H.C.N., Geraldes, R.R., da Silva, F.M.C., Guaita, M.G.D., Camarda, C.M., Szostak, R., Neckel, I.T., Teixeira, V.C., Hesterberg, D., Pérez, C.A., Galante, D., Callefo, F., Neto, A.C.P., Kofukuda, L.M., Sotero, A.P.S., Moreno, G.B.Z.L., Luiz, S.A.L., Bueno, C.S.N.C., Lena, F.R., Westfahl, H., 2023. The CARNAÚBA X-ray nanospectroscopy beamline at the Sirius-LNLS synchrotron light source: Developments, commissioning, and first science at the TARUMÃ station. J Electron Spectros Relat Phenomena 266, 147340. https://doi.org/10.1016/j.elspec.2023.147340

Tripathi, R.P., Chandra, U., Chandra, R., Lokanathan, S., 1978. A Mössbauer study of the effects of heating biotite, phlogopite and vermiculite. Journal of Inorganic and Nuclear Chemistry 40, 1293–1298. https://doi.org/10.1016/0022-1902(78)80037-2

Ulian, G., Valdrè, G., 2023. Crystal-chemical, vibrational and electronic properties of 1M-phlogopite $K(Mg,Fe)_3Si_3AlO_{10}(OH)_2$ from Density Functional Theory simulations. Appl Clay Sci 246, 107166. https://doi.org/10.1016/j.clay.2023.107166

Vasic, B., Czibula, C., Kratzer, M., Neves, B.R.A., Matkovic, A., Teichert, C., 2021. Two-dimensional talc as a van der Waals material for solid lubrication at the nanoscale. Nanotechnology 127, 69–73. https://doi.org/10.1088/1361-6528/abeffe

W. A. Deer, R. A. Howie, and J.Z., 1962. Rock Forming Minerals: Sheet Silicates, 3rd ed. Longmans, London.

Wang, A., Freeman, J.J., Jolliff, B.L., 2015. Understanding the Raman spectral features of phyllosilicates. Journal of Raman Spectroscopy 46, 829–845. https://doi.org/10.1002/jrs.4680

Wei, C., Roy, A., Tripathi, M., Aljarid, A.K.A., Salvage, J.P., Roe, S.M., Arenal, R., Boland, C.S., 2023. Exotic Electronic Properties of 2D Nanosheets Isolated from Liquid Phase Exfoliated Phyllosilicate Minerals. Advanced Materials 35, 2303570. https://doi.org/10.1002/adma.202303570

Woźniak, T., Faria Junior, P.E., Seifert, G., Chaves, A., Kunstmann, J., 2020. Exciton g factors of van der Waals heterostructures from first-principles calculations. Phys Rev B 101, 235408. https://doi.org/10.1103/PhysRevB.101.235408

Zheng, Y.L., Zhao, X.B., Zhao, Q.H., Li, J.C., Zhang, Q.B., 2020. Dielectric properties of hard rock minerals and implications for microwave-assisted rock fracturing. Geomechanics and Geophysics for Geo-Energy and Geo-Resources 6, 22. https://doi.org/10.1007/s40948-020-00147-z

Zhu, K., Pazos, S., Aguirre, F., Shen, Y., Yuan, Y., Zheng, W., Alharbi, O., Villena, M.A., Fang, B., Li, X., Milozzi, A., Farronato, M., Muñoz-Rojo, M., Wang, T., Li, R., Fariborzi, H., Roldan, J.B., Benstetter, G., Zhang, X., Alshareef, H.N., Grasser, T., Wu, H., Ielmini, D., Lanza, M., 2023. Hybrid 2D–CMOS microchips for memristive applications. Nature 618, 57–62. https://doi.org/10.1038/s41586-023-05973-1

Zou, Y.-C., Mogg, L., Clark, N., Bacaksiz, C., Milovanovic, S., Sreepal, V., Hao, G.-P., Wang, Y.-C., Hopkinson, D.G., Gorbachev, R., Shaw, S., Novoselov, K.S., Raveendran-Nair, R., Peeters, F.M., Lozada-Hidalgo, M., Haigh, S.J., 2021. Ion exchange in atomically thin clays and micas. Nat Mater 20, 1677–1682. https://doi.org/10.1038/s41563-021-01072-6